\newcommand{\minpoint}{\mbox{$'\mskip-4.7mu.\mskip0.8mu$}}
\documentstyle[emulateapj]{article}

\lefthead{Fujita et al.}
\righthead{Lyman$\alpha$ Emitters at $z = 3.7$}

\begin{document}


\title{A Search for Lyman$\alpha$ Emitters at Redshift 3.7\altaffilmark{1} }

\author{Shinobu S. Fujita   \altaffilmark{2},
        Masaru Ajiki        \altaffilmark{2},
        Yasuhiro Shioya     \altaffilmark{2},
        Tohru Nagao         \altaffilmark{2},
        Takashi Murayama    \altaffilmark{2},
        Yoshiaki Taniguchi  \altaffilmark{2},
        Sadanori Okamura    \altaffilmark{3,4},
        Masami Ouchi        \altaffilmark{3},
	Kazuhiro Shimasaku  \altaffilmark{3,4},
	Mamoru Doi          \altaffilmark{5},
	Hisanori Furusawa   \altaffilmark{3},
	Masaru Hamabe       \altaffilmark{6},
	Masahiko Kimura     \altaffilmark{7},
	Yutaka Komiyama     \altaffilmark{8},
	Masayuki Miyazaki   \altaffilmark{3},
	Satoshi Miyazaki    \altaffilmark{9},
	Fumiaki Nakata      \altaffilmark{3},
	Maki Sekiguchi      \altaffilmark{7},
	Masafumi Yagi       \altaffilmark{3}, 
	Naoki Yasuda        \altaffilmark{9},
        Yuichi Matsuda      \altaffilmark{10},
        Hajime Tamura       \altaffilmark{10},
        Tomoki Hayashino    \altaffilmark{10},
        Keiichi Kodaira     \altaffilmark{11},
        Hiroshi Karoji      \altaffilmark{9},
        Toru Yamada         \altaffilmark{9},
        Kouji Ohta          \altaffilmark{12}, and
        Masayuki Umemura    \altaffilmark{13}
	}

\altaffiltext{1}{Based on data collected at 
	Subaru Telescope, which is operated by 
	the National Astronomical Observatory of Japan.}
\altaffiltext{2}{Astronomical Institute, Graduate School of Science,
        Tohoku University, Aramaki, Aoba, Sendai 980-8578, Japan}
\altaffiltext{3}{Department of Astronomy, Graduate School of Science,
        University of Tokyo, Tokyo 113-0033, Japan}
\altaffiltext{4}{Research Center for the Early Universe, School of Science,
        University of Tokyo, Tokyo 113-0033, Japan}
\altaffiltext{5}{Institute of Astronomy, Graduate School of Science, 
        University of Tokyo, Mitaka, Tokyo 181-0015, Japan}
\altaffiltext{6}{Department of Mathematical and Physical Sciences,
	Faculty of Science, Japan Women's University, Tokyo 112-8681, Japan}
\altaffiltext{7}{Institute for Cosmic Ray Research, 
	University of Tokyo, Kashiwa, Chiba 277-8582, Japan}
\altaffiltext{8}{Subaru Telescope, National Astronomical Observatory, 
	650 N.A'ohoku Place, Hilo, HI 96720}
\altaffiltext{9}{National Astronomical Observatory, 
	Mitaka, Tokyo 181-8588, Japan}
\altaffiltext{10}{Research Center for Neutrino Science, 
        Graduate School of Science,
        Tohoku University, Aramaki, Aoba, Sendai 980-8578, Japan}
\altaffiltext{11}{The Graduate University for Advanced Studies,
        Shonan Village, Hayama, Kanagawa 240-0193, Japan}
\altaffiltext{12}{Department of Astronomy, Graduate School of Science, 
        Kyoto University, Kitashirakawa, Sakyo, Kyoto 606-8502, Japan}
\altaffiltext{13}{Center for Computational Physics, University of Tsukuba,
        1-1-1 Tennodai, Tsukuba 305-8571, Japan}

\begin{abstract}

We present the results of a survey for emission-line objects based on
optical intermediate-band 
($\lambda_{\rm c}$ = 5736 \AA ~ and $\Delta\lambda$ = 280 \AA)
and broad-band ($B$, $V$, $R$, and $i^\prime$)  observations of the 
Subaru/XMM-Newton Deep Field on the 8.2 m Subaru telescope with
the Subaru Prime Focus Camera, Suprime-Cam.
All the data were obtained during 
the guaranteed time observations of the Suprime-Cam instrument.
The intermediate-band image covered a sky area with
10\minpoint62 $\times 12\minpoint40 \approx 132$ arcmin$^2$
in the Subaru/XMM-Newton Deep Field (Ouchi et al.).

Using this image, we have found 23 emission-line sources whose
observed emission-line equivalent widths are greater than 250 \AA.
Their optical multicolor properties indicate that
six emission-line sources are
Ly$\alpha$ emitters at $z \approx$ 3.7 ($\Delta z \approx 0.22$).
They are either intense starburst
galaxies or active galactic nuclei like quasars at $z \approx$ 3.7.
Two more emission-line sources may also be Ly$\alpha$ emitters at 
$z \approx$ 3.7 although their multicolor properties are marginal.
Among the remaining 15 emission-line objects, eight objects
appear strong emission-line galaxies at lower redshift; e.g., [O {\sc ii}]
$\lambda$3727 emitters at $z \approx 0.54$, H$\beta$
at $z \approx 0.18$, or
[O {\sc iii}]$\lambda$5007 emitters at $z \approx 0.15$.
The remaining seven objects are unclassified because they are too
faint to be detected in broad-band images.
We discuss observational properties of these strong emission-line sources.
In particular, our data allow us to estimate the star formation density
at $z \approx 3.7$ for the first time.

\end{abstract}


\section{INTRODUCTION}

It has been often argued that
forming galaxies at high redshifts experienced very luminous
starbursts and thus they could be very bright in line emission such as
Ly$\alpha$ and [O {\sc ii}]$\lambda$3727 emission lines
(e.g., Partridge \& Peebles 1967; Larson 1974; Meier 1976).
However, although many attempts were made to search for such
very strong emission-line sources at high redshifts
(e.g., Pritchet 1994; see also Pahre \& Djorgovski 1995;
Thompson, Mannucci, \& Beckwith 1996),
most these searches by mid 1990's failed except some successful
surveys around known high-$z$ objects such as quasars and radio galaxies 
(Hu \& McMahon 1996; Hu, McMahon,
\& Egami 1996; Petitjean et al. 1996; Pascarelle et al. 1996; Keel et al. 1999).
Therefore, the Lyman break method (or the broad-band, color-selection method)
has been mainly used to investigate observational properties of high-$z$ 
galaxies for these past several years
(Steidel et al. 1996a, 1996b, 1999; Cowie et al. 1996; 
Lanzetta, Yahil, \& Fernandez-Soto 1996;
Madau et al. 1996; Yahata et al. 2000;  Ouchi et al. 2001).

Recently, however, new attempts with 10-m class telescopes have revealed
the presence of Ly$\alpha$ emitters in blank fields at high redshift
(Cowie \& Hu 1998, hereafter CH98; see also Hu, McMahon, \& Cowie 1998, 1999; 
Ouchi et al. 2002; Hu et al. 2002). 
Subsequently, Steidel et al. (2000; hereafter S00) also succeeded in
finding a number of high-$z$  Ly$\alpha$ emitters in the SSA22 blank field.
Furthermore, Kudritzki et al. (2000; hereafter K00) 
have identified nine Ly$\alpha$ emitters at $z \approx 3.1$ during the 
course of their narrow-band imaging survey aimed at 
looking for intracluster planetary nebulae in the Virgo cluster 
(Mendez et al. 1997).
These surveys have reinforced the potential importance of
search for high-$z$ Ly$\alpha$ emitters. Thus, deep imaging surveys
with narrow-band filters are now considered to be a powerful tool
 in this 10m-class telescopes era
to probe the Ly$\alpha$ emission from high-$z$ young galaxies
and active galactic nuclei like quasars.
Such surveys for high-$z$ emission-line sources provide us with very important 
information not only on the formation and evolution of galaxies
but also on the cosmic reionization process at very high redshift
(e.g., Loeb \& Barkana 2001 and references therein). 

However, such surveys for emission-line galaxies with narrow-band filters
have the limitation that survey volumes are so small because of 
narrower band widths (e.g., $\sim 100$ \AA). In order to gain survey volumes,
we need very wide-field CCD cameras on 10-m class telescopes. Fortunately, 
the Subaru 8.2 m telescope (Kaifu 1998) at Mauna Kea Observatories 
has the wide-field (a $34^\prime \times 27^\prime$ field of view)
prime focus camera,
Suprime-Cam (Miyazaki et al. 1998). This camera enables us to carry out 
narrow-band imaging surveys for high-$z$ emission-line objects\footnote{
Wide-field CCD cameras are also available on some 4-m class telescopes.
One is the wide-field camera (a 36$^\prime \times 36^\prime$ field of view)
on the Kitt Peak National Observatory's 4m Mayall telescope.
Together with narrow-band filters, this camera has been 
used to search for high-$z$ Ly$\alpha$ emitters;
the Large Area Lyman Alpha (LALA) survey [Rhoads et al. 2000;
Rhoads \& Malhotra 2001; Malhotra \& Rhoads 2002 (hereafter MR02)].
The other is the Taurus Tunable Filter system on the 3.9 m AAT telescope
(Bland-Hawthorn \& Jones 1997; Baker et al. 2001). Although the current
field of view is $\approx 10^\prime$ in diameter, this tunable filter
system will have a field of view of  $\approx 30^\prime$ in diameter if
available on the prime focus of some 4-m or 10-m class telescopes
(Bland-Hawthorn et al. 2001).}.

For this purpose, we made a new filter system which consists of 20 
filters with the spectral resolution of $R=23$ covering from 4000 \AA ~ to 
9500 \AA ~ (Hayashino et al. 2000; Taniguchi 2001; Shioya et al. 2002). 
Its spectral resolution is not higher than those of typical narrow-band 
filters used for Ly$\alpha$ emitter searches; e.g.,  $R=70$ with the central 
wavelength $\lambda_{\rm c}$ = 5390 \AA~
and the band width of $\Delta\lambda$ = 77 \AA ~ (S00), and 
$R=62$ with $\lambda_{\rm c}$ = 4970 \AA ~ and $\Delta\lambda$ = 80 \AA ~ 
(S00). 
However, our intermediate-band filter system (called as the IA system which
means the intermediate-band filter set A) is useful to detect strong 
emission-line sources whose emission-line equivalent widths exceed 
250 \AA ~ in the observed frame. Further, our system covers the entire 
optical wavelength range and thus we will be able to carry out any systematic 
searches for strong emission-line sources at various redshifts from 
$z \approx 2.2$ to $z \approx 6.9$ (Taniguchi 2001).

During the commissioning phase of Suprime-Cam on the Subaru telescope,
we made an imaging survey for Ly$\alpha$ emitters at $z \approx 3.7$
in the Subaru/XMM-Newton filed (Ouchi et al. 2001)
using one of IA filters (IA574; 
$\lambda_{\rm c}$ = 5736 \AA ~ and $\Delta\lambda$ = 280 \AA).
In this paper, we present our first results of this intermediate-band
imaging survey.

Throughout this paper, magnitudes are given in the AB system.
We adopt a flat universe with $\Omega_{\rm matter} = 0.3$, 
$\Omega_{\Lambda} = 0.7$ 
and $h=0.7$ where $h = H_0/($100 km s$^{-1}$ Mpc$^{-1}$).

\section{OBSERVATIONS AND DATA REDUCTION}

\subsection{Observations}

As described in Ouchi et al. (2001; see also Ouchi 2001), we have obtained
deep and wide-field $B$-, $V$-, $R$-, and $i'$-band imaging data 
of a central $30'\times 24'$ area in the 
Subaru/XMM-Newton Deep Survey Field centered at
$\alpha$(J2000) = $2^{\rm h} 18^{\rm m} 00^{\rm s}$ and
$\delta$(J2000) = $-5^\circ 12 ' 00''$ using 
Suprime-Cam (Miyazaki et al. 1998) 
on the Subaru telescope during a period between 2000 August and
2000 November.  
In addition to these broad-band image data, we have obtained 
an intermediate-band image using the IA filter IA574
($\lambda_{\rm c}$ = 5736 \AA ~ and $\Delta\lambda$ = 280 \AA)
in 2000 August. The transmission curves of the filters 
used in our observations are shown in Figure 1.

During the IA574-band observing run, only 4 CCD chips, each of which has
2048$\times$4096 pixels, were installed in the Suprime-Cam.
Two CCD chips are MIT ones while the others are SITe ones;
a north-east part (13\minpoint7 $\times 13\minpoint7$)
of the broad-band image was observed with the MIT chips and 
a north-west part (13\minpoint7 $\times 13\minpoint7$)
was observed with the SITe ones. 
In this paper, we present results obtained with the IA574 image
taken with the SITe CCDs.
The final sky coverage in our analysis is 
10\minpoint62 $\times 12\minpoint40$ in the 13\minpoint7 $\times 13\minpoint7$
area covered by the SITe CCD chips.
Our sky coverage is shown 
together with that of the broad-band imaging (Ouchi et al. 2001) 
in Figure 2.
Results  obtained with the IA574
image with the MIT CCDs will be given in Yoshida et al. (2002).
A journal of our all observations is summarized in Table 1.

The individual CCD data were reduced and
combined using IRAF and the mosaic-CCD data reduction software 
developed by us (Yagi 1998).
The following photometric standard stars were observed to 
calibrate the data; 1) PG 0205+134 (Massey et al. 1988) and G158$-$100
(Oke 1990) for the August run, and 2) SA92, SA98, SA 95, SA101
(Landolt 1992), and SA95\_42 (Oke 1990).
The combined images for the individual bands were aligned and 
smoothed with Gaussian kernels to match their seeing sizes. 
The final images cover a contiguous
618 arcmin$^2$ area with a PSF FWHM of $0.''98$ for the broad-band
data while 188 arcmin$^2$ area with a PSF FWHM of $0.''98$ for the IA574 data.
In the later analysis, we use the 132 arcmin$^2$ area covered by 
both the IA574 and the broad-band data. 
The final IA574 image is shown in Figure 3 together with a final
color image made by using the broad-band images. 

The total size of the field is 10\minpoint62 by 12\minpoint40, corresponding 
to a total solid angle of $\approx$ 132 arcmin$^{2}$. The effective volume 
probed by the IA574 imaging has (co--moving) transverse dimensions of 21.5 
$h_{0.7}^{-1}\times 25.1 h_{0.7}^{-1}$ Mpc, and the half--power points of 
the filter correspond to a
co--moving depth along the line of sight of 173.8 $h_{0.7}^{-1}$ Mpc
($z_{\rm min} \approx 3.60$ and $z_{\rm max} \approx 3.83$), 
where $h_{0.7}=H_0/(70 {\rm km \; s^{-1} \; Mpc^{-1}})$.
Therefore, a total volume
of  93,952 $h_{0.7}^{-3}$ Mpc$^{3}$ is probed in our IA574 image, being much 
wider than
those of the previous pioneering studies, 
10,410  $h_{0.7}^{-3}$ Mpc$^{3}$ (CH98) and
16,741 $h_{0.7}^{-3}$ Mpc$^{3}$ (S00) (see Table 6).

\subsection{Source Detection and Photometry}

Source detection and photometry were performed using
SExtractor version 2.1.6 (Bertin \& Arnouts 1996).
Here we used the same source-detection criterion as that 
in Ouchi et al. (2001); a source must be a 5-pixel connection
above 2$\sigma$.  The detection limit of the IA574 image is
25.7 mag for a 3$\sigma$ detection with a
$2''$ diameter aperture.

As for the source detection in the broad-band images,
the limiting magnitudes are $B=27.6$, $V=26.4$, $R=26.7$, and
$i'=26.2$ for a 3$\sigma$ detection with a
$2''$ diameter aperture (Ouchi et al. 2001).

In the above source detection, we have detected $\sim 5500$ sources
down to $IA574 = 26$.
In order to examine the completeness in the IA574 imaging,
we show results of the number count as a function of IA574 magnitude
in Figure 4. This figure shows that our source detection appears
complete down to $IA574 \approx$ 25.4. 

In order to make sure that our source selections were done with little
ambiguity, we have newly performed a simulation using the IRAF ARTDATA
(e.g., Kajisawa et al. 2000).
We assume that galaxies have two types of light distributions obeying
a) the de Vaucouleurs' $r^{1/4}$ law and b) the exponential law.
For each type of galaxies, we generated 300 model galaxies for each
magnitude interval ($\Delta m$ = 0.2 mag); i.e., 600 model galaxies in total.
Their sky positions, half-right radius (1 kpc to 7 kpc), and ellipticities
are randomly determined. Then these model galaxies are put into the CCD
data together with Poisson noises. After smoothing model-galaxy images
to match to the seeing size, we try to detect them using SExtractor with
the same procedure as that used in our paper.
The detectability of the model galaxies in each band is shown in
Figure 5 as a function of AB magnitude.
As for objects brighter than $IA574 = 25.4$, we find that the
detectability is higher than 50\%. Therefore, we consider that
this result appears to be consistent with the completeness limit,
$m_{\rm lim}$(AB) = 25.4  
shown in Figure 4 (see also the last column in Table 1).

The $VR - IA574$ color is plotted for the detected model galaxies 
with a color of $VR-IA574=0$ 
as a function of IA574 magnitude in Figure 6.
It is shown that almost all the model galaxies are within
2 sigma deviations.
Therefore, we conclude that our source selections were done
with appropriate accuracy for our purpose.

\section{RESULTS}

\subsection{Selection of IA574-Excess Objects}

Since the central wavelength of the $V$ filter is bluer than 
that of the IA574 filter (5736 \AA),
we constructed an image that we will refer to as the ``VR continuum'' using
a linear combination ($VR=3.4V+1.0R$) 
of the deep V and R images after scaling them to
the same photometric zero point; a 3 $\sigma$ photometric limit of
$VR \simeq 26.6$ in a 2 arcsec diameter aperture.
This enables us to more precisely sample the continuum at the same effective
wavelength as that of the IA574 filter.

Although the detection limit of the IA574 image is 25.7,
our source detection in IA574 appears complete down to $\approx$ 25.4
from Figure 4. Therefore, 
we have tried to detect IA574-excess objects
down to $IA574 = 25.4$ and made an IA574-selected catalog
in which 3635 objects are contained.
In Figure 7, we show the diagram between $VR-IA574$ and
$IA574$ for the objects in the above catalog.
Taking the scatter in the $VR-IA574$ color into account, we have selected
strong emission-line sources with the criteria of 
$VR-IA574 \geq 0.7$ and $IA574 \leq 25.4$.
These criteria are shown by dotted lines in Figure 7.
There are 101 sources which satisfy the above two criteria.
We also show the distributions of 2$\sigma$ (solid lines)
and 3$\sigma$ (dashed lines) errors in Figure 7.
We remove 20 objects out of 101 sources because their
$VR-IA574$ colors are smaller than the 2$\sigma$ error.

Since the central wavelength of the IA574 filter is 
closer to that of the V-band filter than to that of the
R-band one, we have adopted the criterion of $VR-IA574 \geq 0.7$ 
as our primary criterion. However, red-color objects with 
a continuum break in the V-band window may also be detected
as strong emission-line sources even though they have little
emission-line flux. In order to remove such objects,
we have also adopted another criterion of $R-IA574 \geq 0.7$.
In Figure 8, we show the diagram between $VR-IA574$ and $R-IA574$
for the 81 objects found with the $VR-IA574$ color selection.
As shown in this figure, 32 sources have been rejected because
they do not show significant excess in the $R-IA574$ colors.
Then, we obtain a sample of 49 objects with both 
$VR-IA574 \geq 0.7$ and $R-IA574 \geq 0.7$.
These color criteria mean that all the sources have 
their emission-line equivalent widths higher than 250\AA.
It is also noted that dusty starburst galaxies with very red colors
may not be found with our selection criteria and thus 
the strong emission-line objects found in our search may be  
mostly either starburst galaxies with weak reddening  
or typical type 1 AGNs such as quasars.

In order to secure that our selection is reliable,
we adopt another severe criterion; a source must be
a 13-pixel connection above 2$\sigma$. The reason for this
is that a source under a $\approx$ 1 arcsec seeing condition
has a 13-pixel connection. Applying this criterion,
we reject 24 sources among the 49 sources.
Finally, we have made visual inspection of all candidates' images 
carefully in order to reject ambiguous objects which may 
be attributed to noises. In this procedure, we have
rejected two sources because they show a linear or an unusual shape.
Then we obtain our final sample of 23 emission-line sources.

We have also checked any of our line emitter candidates are 
neither variable objects nor moving objects 
by comparing the $R$-band image obtained in 2000 August with 
that in 2000 November. 
We do not find any spatially-extended Ly$\alpha$ emitters like
Ly$\alpha$ blobs found by S00. See for Ly$\alpha$ blobs
Taniguchi \& Shioya (2000), and Taniguchi, Shioya, \& Kakazu (2001)
and references therein.

\subsection{Selection of IA574-Excess Objects at z$\approx$3.7}

Our main aim in the present survey is to find strong Ly$\alpha$ emitters
at $z \approx 3.7$. However, strong emission-line sources at lower redshift
may be also found in our survey;
e.g., C {\sc iv} $\lambda$1550 sources at $z \approx 2.70$,
Mg {\sc ii} $\lambda$2798 sources at $z \approx 1.05$,
[O {\sc ii}]$\lambda$3727 sources at $z \approx 0.54$,
H$\beta ~ \lambda$4861 at $z \approx 0.18$,
[O {\sc iii}]$\lambda$5007 sources at $z \approx 0.15$, and so on.
In order to distinguish Ly$\alpha$ emitters at $z \approx 3.7$ from
emission-line objects at lower redshift,
we investigate their broad-band color properties.
In this procedure, we also take account of the observed
emission-line equivalent widths.

First, we show that the $B-R$ color provides a nice tool to 
pick up Ly$\alpha$ emitters at $z \approx 3.7$.
In Figure 9, we show the diagram of $B-R$ color as a function of 
redshift for galaxies with spectral energy distributions (SEDs) typical to
E (the bulges of M31 and M81), Sbc, Scd, and Irr 
(Coleman, Wu, \& Weedman 1980; hereafter CWW).
The CWW's SEDs cover a wavelength range from 1500 \AA ~ to 10000 \AA.  
We therefore extend them below 1500 \AA ~ assuming 
$f_{\lambda} \propto \lambda^{-0.82}$ (Kinney et al. 1993) 
down to 912 \AA.  
As an SED of young starburst galaxies, we adopt an SED generated
by the population synthesis model GISSEL96 (Bruzual \& Charlot 1993);
a galaxy with a constant star formation rate at an age of 10$^8$ yr.
We also show expected redshift ranges for Ly$\alpha$ emitters at 
$z \approx 3.7$, [O {\sc ii}]$\lambda$3727 emitters at
$z \approx 0.54$, and [O {\sc iii}]$\lambda$5007 emitters 
at $z \approx 0.15$.

For low-$z$ galaxies, their $B-R$ colors are mainly determined
by the stellar populations. On the other hand, for high-$z$
galaxies beyond $z \sim 2.5$, their colors are mainly determined
by the continuum depression due to the intergalactic extinction;
i.e., cosmic transmission (e.g., Madau 1995).
In Figure 9, we show three cases of different cosmic transmissions;
1) the mean value of the cosmic transmission by Madau (1995)
(solid curves),
2) twice as large as the above mean value (dotted lines), and
3) a half of the above mean value (dashed lines). 
These results imply that Ly$\alpha$ emitters at $z \approx 3.7$
have $B-R \gtrsim 1.0$ even if the cosmic transmission 
shows scatters within a factor of two from one line of sight to
another.

It is also expected that E-,
Sbc-, and Scd-type galaxies at lower  
redshift, have $B-R > 1$. However, our strong emission-line
objects found in this study have large emission-line equivalent
widths; i.e., $EW_{\rm obs} \geq$ 250\AA. This value corresponds
to a rest-frame equivalent width of $EW_0$=163\AA ~ for 
[O {\sc ii}]$\lambda$3727 emitters at $z \approx 0.54$, 
$EW_0$=212\AA ~ for H$\beta$ emitters at $z \approx 0.18$, and
$EW_0$=217\AA ~ for  [O {\sc iii}]$\lambda$5007 emitters
at $z \approx 0.15$.
Since it is known that typical rest-frame [O {\sc ii}] or
[O {\sc iii}] emission-line galaxies in nearby universe
have  $EW_0 < 100$\AA ~ (e.g., Jansen et al. 2000),
it is unlikely that the strong emission-line sources
with $B - R \geq 1.0$ are low-$z$ sources (see also section 3.3).

Second, we show the diagram between $B-R$ and $R-i^\prime$
for all the objects detected in the broad-band $B$, $R$, and 
$i^\prime$ images in the right panel of Figure 10. 
We also show color evolutions of
model galaxies with the CWW SEDs typical to 
E, Sbc, Scd, and Irr and with the SED for young starburst 
as a function of redshift (left panel).
The model results show that Ly$\alpha$ emitters at $z \sim 3.7$
may occupy the shaded domain defined with both $B-R > 1.0$
and $R-i^\prime \lesssim 0.7$ although 
the latter color constraint appears not so strong.
This figure also implies again that either strong [O {\sc ii}]
or [O {\sc iii}] emitters could have $B-R < 1$.

It seems necessary to examine whether or not 
strong C {\sc iv} emitters (i.e., quasars) at $z \approx 2.7$ 
are misclassified as Ly$\alpha$ emitters at $z \approx 3.7$. 
Using the composite spectrum of SDSS quasars (Vanden Berk et al.
2001), we estimate the $B-R$ color when we observe a quasar 
with the SDSS composite spectrum at  $z \approx 2.7$.
We find that its $B-R$ color is much bluer than 1.0
because the Lyman break comes shorter than the $B$-band
transmission. Therefore, there is no possibility to select
quasars at $z \approx 2.7$ when we use the above color
criteria.

In conclusion, one can identify Ly$\alpha$ emitters
at $z \approx 3.7$ using the two color criteria;
1) $B-R \geq 1.0$, and 2) $R-i^\prime \leq 0.7$.
It is again noted that the observed larger emission-line
equivalent widths (i.e., $EW_{\rm obs} \geq$250\AA) 
allow us to adopt the above simple color criteria
in our selection.
In this way, we have classified our 23 emission-line sources
into the following four categories.

(1) Ly$\alpha$ emitters at $z \approx 3.7$:
Six emission-line objects satisfy the above two criteria and
thus they are identified as  Ly$\alpha$ emitters at $z \approx 3.7$
which are marked by open red circles in the upper shaded region in Figure 10. 
Their positions, emission-line equivalent widths, and 
photometric properties  are given in Table 2.
In Figure 11, we show the $B$, $V$, $IA574$, $R$, and $i^\prime$ 
images of each object in our sample of Ly$\alpha$ emitters 
at $z \approx 3.7$. The SED is also shown in the right panel
for each object.
We note that one object (No. 4) appears to be extended in the the IA574 image.
Its angular diameter (above 2 sigma) is estimated as 2.0 arcsec,
corresponding to the linear diameter of 14.3 $h_{0.7}^{-1}$ kpc at $z=3.7$.
Although we cannot rule out the possibility that this source
is a low-z object, we do not adopt any criterion on the source size
in our source selection procedure. Therefore, we include this source
as a Ly$\alpha$ emitter candidate.

(2) Marginal Ly$\alpha$ emitters at $z \approx 3.7$: 
Two objects shown by pink colors in Figure 10 appear marginal
between objects at $z \approx 3.7$ and ones at lower redshift
because their $B - R$ colors marginally satisfy the condition
of $B - R \geq 1.0$ if their observational errors are taken
into account. Therefore, we call them ^^ ^^ marginal"
Ly$\alpha$ emitters at $z \approx 3.7$.
Their positions, emission-line equivalent widths, and
photometric properties  are given in Table 3.
In Figure 12, we show the $B$, $V$, $IA574$, $R$, and $i^\prime$
images of each object in our sample of marginal Ly$\alpha$ emitters
at $z \approx 3.7$. The SED is also shown in the right panel
for each object.

(3) Low-$z$ emission-line objects:
Eight objects among the remaining 15 objects which are marked 
by open squares in Figure 10
may be emission-line objects at lower redshifts because their 
$B - R$ colors are significantly bluer than 1.0.
Their positions, emission-line equivalent widths, and 
photometric properties are given in Table 4.
In Figure 13, we show the $B$, $V$, $IA574$, $R$, and $i^\prime$ 
images of each object in our sample of low-$z$ emitter candidates.
The SED is also shown in the right panel for each object.

(4) Unclassified emission-line objects:
The six objects out of the remaining seven are detected only in the IA574.
Therefore, we cannot use their broad-band photometric properties 
and thus they are remained as ^^ ^^ unclassified" emission-line
objects.
The last one is detected only in $R$ image except for IA574.
Their positions, emission-line equivalent widths, and 
photometric properties are given in Table 5.
In Figure 14, we show the $B$, $V$, $IA574$, $R$, and $i^\prime$ 
images of each object.
The SED is also shown in the right panel for each object.

\subsection{Equivalent Widths of Emission Features}

First, we investigate the emission-line equivalent width of 
the low-$z$ emission-line sources.
We show the rest-frame [O {\sc ii}] and [O {\sc iii}] emission-line
equivalent widths (crosses) as a function of $B-R$ in Figures 15 and 16,
respectively. Note that we use the $B-R$ color in the Vega-based
photometric system,  $(B-R)_{\rm Vega}$
(e.g., Fukugita, Shimasaku, \& Ichikawa 1995)
in order to compare the observations with model results which are
obtained by using the population synthesis model
GISSEL96 (Bruzual \& Charlot 1993). In our model calculations,
we use the $\tau$ model with both $\tau$ = 1 Gyr and
the metallicity of $Z=0.02$.
Model results for the $(B-R)_{\rm Vega}$ are obtained for the following ages;
10, 100, 500 Myr, 1, 2, 3, 4, 7, and 10 Gyr.
For each model, we derive the H$\beta$ luminosity from the Lyman continuum
luminosity using the following formula (Leitherer \& Heckman 1995),
\begin{equation}
L({\rm H}\beta) = 4.76 \times 10^{-13} N({\rm H}^0) ~ {\rm ergs \; s^{-1}},
\end{equation}
where $N({\rm H}^0)$ is the number density of Lyman continuum photon
in units of s$^{-1}$.
The equivalent widths of [O {\sc ii}] and [O {\sc iii}] emission
are estimated for the following cases;
1) log [O {\sc ii}]/H$\beta$ = 0 (the lower solid line in Figure 15)
and 0.5 (the upper solid line in Figure 15), and
2) log [O {\sc iii}]/H$\beta$ = $-0.5$ (the lower solid line in Figure 16)
and 0.5 (the upper solid line in Figure 16).
The low-$z$ emission-line candidates found in our survey appear to show
much stronger emission-line galaxies than star-forming galaxies
found in the local universe. Such examples have been indeed found by
Ohyama et al. (1999); e. g., [O {\sc ii}] emitters
at $z \sim 0.5$ have $EW_{\rm obs} \sim 200$\AA.
Such galaxies must be very blue
and thus their $B-R$ colors are expected to be much bluer than
$B-R$=1 (see also Stockton \& Ridgeway 1998; Stern et al. 2000). 
These results also reinforce that that low-$z$ strong emission-line galaxies
do not have $B-R > 1$.

Second, we compare the distribution of observed emission-line
equivalent widths ($EW_{\rm obs}$) among the four samples in Figure 17,
(1) the Ly$\alpha$ emitter sample, (2) the marginal Ly$\alpha$ emitter sample,
(3) the low-$z$ emitter sample, and (4) the unclassified sample.
We obtain the average equivalent widths of
482$\pm$246\AA ~ for the Ly$\alpha$ emitter sample.
If we combine the Ly$\alpha$ emitter sample and the 
marginal ones, we obtain the average equivalent widths of
575$\pm$280\AA. On the other hand, we obtain the average equivalent 
widths of for the  low-$z$ sample, being smaller than the 
above values. This makes sense because $EW_{\rm obs}$ of the Ly$\alpha$ emitter
candidates is amplified by a factor of $\approx$4.7 while the
amplification factor must be rather small for the low-$z$ emitter candidates;
e.g., a factor of 1.54 for [O {\sc ii}] emitters.

Finally, it is interesting to note again that the low-$z$ emitters detected
in our survey may have large rest-frame emission-line equivalent widths;
e.g., $EW_0 \sim$ 150 -- 200 \AA ~ if they are either, [O {\sc ii}],
H$\beta$, or [O {\sc iii}] emitters. As noted before, 
it is known that typical rest-frame [O {\sc ii}] or
[O {\sc iii}] emission-line galaxies in nearby universe have  $EW_0 < 100$\AA ~
(e.g., Jansen et al. 2000).
If emission-line galaxies are located at $z \approx 0.151$ -- 0.174, 
both H$\beta$ and [O {\sc iii}]$\lambda\lambda$4959,5007 emission lines
can be detected simultaneously in our $IA574$ image, resulting in 
larger-than-normal emission-line equivalent widths. However,
it seems unlikely that most of the low-$z$ emitters are located
at the above narrow redshift range. Furthermore, 
Ohyama et al. (1999) found very strong [O {\sc ii}] emitters
at $z \approx 0.5$ serendipitously.
In Figure 18, we show a diagram between $EW_{\rm obs}$ and $VR$
for the eight low-$z$ emitters. It appears that fainter galaxies
tend to have larger $EW_{\rm obs}$.
Therefore, it is suggested that
a number of emission-line galaxies with large equivalent widths 
may have not yet been probed in previous surveys because they are
too faint to be included in bright magnitude-limited samples. 

\subsection{Spatial Distribution of the IA574-Excess Objects}

We investigate the spatial distributions of the Ly$\alpha$
emitter candidates at $z \approx 3.7$. In Figure 19, we plot the spatial
distributions for the four emitter samples.
The Ly$\alpha$ emitter candidates at $z \approx 3.7$ are shown by
open circles. Five among the six candidates together with 
the two marginal  Ly$\alpha$ emitter candidates are distributed in
the southern part of our image. However, the low-$z$ emitter sample
also shows such a tendency.  
We do not discuss further detail on this issue
because our survey depth is not so deep and spectroscopic confirmation
has not yet been done.

\section{DISCUSSION}

\subsection{Nature of the Ly$\alpha$ Emitter Candidates at $z \approx 3.7$}

We have detected the six and two marginal candidates of Ly$\alpha$ 
emitters at $z \approx 3.7$.
Although it is uncertain whether they are genuine star-forming galaxies 
or active galactic nuclei, it is highly probable that the emission feature is
attributed to  Ly$\alpha$ emission and thus  their redshifts are $z 
\approx 3.7$. Therefore, it seems interesting to investigate their rest-frame
Ly$\alpha$ equivalent widths, $EW_0$(Ly$\alpha$) (see Table 7). 
We note that because of the effect of cosmic transmission, the value of $EW_0$(Ly$\alpha$) 
evaluated here is smaller than the intrinsic value (see appendix). 
In Figure 20, we show a histogram of $EW_0$(Ly$\alpha$) for the six sources.
It is shown that the rest-frame equivalent widths range from 57\AA ~ to 
216\AA. The average value is  
$<EW_0$(Ly$\alpha)> \simeq 103$ \AA ~ $\pm$ 53 \AA.
These values are comparable to those 
(except the two Ly$\alpha$ blobs found by S00)
found by  CH98 and S00. 
It is noted that Vanden Berk et al. (2001) obtained 
$<EW_0$(Ly$\alpha)> \simeq 93$ \AA ~ $\pm$ 0.7 \AA ~
for a sample of over 220 quasars found in the Sloan 
Digital Sky Survey. This median value is also similar to
the median value obtained for our sample.

In Figure 21, we compare our result with those of the previous
narrow-band surveys by CH98, S00, and MR02. Since their survey volumes
are different from ours, we have re-evaluated 
the frequency distributions of the equivalent widths so as to
match to our survey volume. For the results by MR02, we adopt the
frequency distribution in which 1$\sigma$ $R$-band continuum is used
in the estimate of equivalent widths. As shown in this figure, 
all the previous surveys have detected more numerous Ly$\alpha$ 
emitters by a factor of 3 to 9 than our survey for objects
with $EW_0$ = 50 to 100 \AA. The reason for this seems that 
their survey depths in $EW$ are deeper than that of our survey.

Both our survey and the LALA survey by MR02 have succeeded in detecting
stronger Ly$\alpha$ emitters with $EW_0 > 100$ \AA.
Since it is considered that such strong emitters may be rarer than
weak emitters as usual, their wide-field coverages enable them to detect
such strong emitters. 
We also note that the LALA survey
have detected more numerous sources whose equivalent widths reach $\sim$
500 \AA. This is because their survey depths in flux are deeper than ours.  

In Figure 22, we show a diagram between $EW_0$(Ly$\alpha$) and $VR$ magnitude
for the six sources. It is found that the fainter objects tend to
have larger $EW_0$(Ly$\alpha$). This tendency can be understood in terms of
the so-called Baldwin effect (Baldwin 1977) for active galactic nuclei.
However, it is also known that such tendency can be found for star-forming
galaxies (e.g., Cowie et al. 1996) although the correlation shows much larger
scatters than that for active galactic nuclei.
Since the correlation shown in Figure 22 exhibits the large scatter,
it is suggested that the majority of the detected Ly$\alpha$ emitters
are star-forming galaxies like those found by CH98 and S00
rather than quasars.

\subsection{Space Density of the Ly$\alpha$ emitters at $z \approx 3.7$}

We have detected the 6 candidates of Ly$\alpha$ emitters at $z \approx 3.7$
in the volume of  93,952 Mpc$^3$. This gives a space density of the Ly$\alpha$ 
emitters, $n({\rm Ly}\alpha) \simeq 6.4 \times 10^{-5}$ Mpc$^{-3}$. 
CH98 obtained  $n({\rm Ly}\alpha) \simeq 9.6 \times 10^{-4}$ Mpc$^{-3}$ for 
the Hubble Deep Field and  
$n({\rm Ly}\alpha) \simeq 1.3 \times 10^{-3}$ Mpc$^{-3}$
for the SSA22 field. S00 obtained 
$n({\rm Ly}\alpha) \simeq 4.3 \times 10^{-3}$ Mpc$^{-3}$
for the LBG (Lyman break galaxies)  overdensity region. 
Further, K00 obtained 
$n({\rm Ly}\alpha) \simeq 1.3 \times 10^{-3}$ Mpc$^{-3}$
in their La Palma field (Mendez et al. 1997).
The density we obtained is lower by one order of magnitude than their values.
This difference may be partly due to that their survey depths are 
deeper by a factor of 3 than that of our survey. 
However, the higher density obtained by S00 may be real
because their survey field is the LBG overdensity region.
A summary of the space densities of high-$z$ Ly$\alpha$ emitters 
is given in Table 6.

Then we investigate the Ly$\alpha$ luminosities of the $z \approx 3.7$ 
candidates. We assume that all the sources have a redshift of 3.717
which corresponds to the case that the Ly$\alpha$ is shifted to
the central wavelength of IA574 filter.
In Table 7, we give the Ly$\alpha$ luminosities for the 6 objects.
The derived  Ly$\alpha$ luminosities range from 
$\approx 5 \times 10^{42}$ to $1 \times 10^{43}$ ergs s$^{-1}$,
being slightly larger than those of CH98's sources.
In Figure 23, we show the distribution of 
Ly$\alpha$ luminosities for our sample together with the results of
CH98. Here we estimate the Ly$\alpha$ luminosities of
CH98's sources using the cosmology adopted in this paper.
It is shown that our survey probes higher-luminosity sources 
with respect to the CH98 survey. 
It is likely that higher-luminosity sources are fewer than 
lower-luminosity ones. Therefore, in order to find such higher-luminosity
sources, it is necessary to perform wider-field surveys. 
Since our survey volume is wider by a factor of 10 than that of CH98,
we can detect such higher-luminosity Ly$\alpha$ emitters
in our survey.
On the other hand, our survey limit ($EW_{\rm limit} = 250$\AA)
is shallower by a factor of 2.5 than their limit 
($EW_{\rm limit} \approx 100$\AA).
Therefore, we miss a large number of lower-luminosity Ly$\alpha$ emitters.
Wide-field and very deep narrow-band imaging surveys will be necessary to
explore the nature of emission-line objects at high redshifts
(e.g., Taniguchi et al. 2001). 

\subsection{Star Formation Density at $z \approx 3.7$}

Finally we estimate the star formation rate for the six Ly$\alpha$ 
emitters at $z \approx 3.7$. 
Given the formula,

\begin{equation}
SFR = 7.9 \times 10^{-42} L({\rm H}\alpha) ~~~ M_\odot {\rm yr}^{-1}
\end{equation}
where $L({\rm H}\alpha)$ is the H$\alpha$ luminosity in units of
ergs s$^{-1}$ (Kennicutt 1998)  together with a relation

\begin{equation}
L({\rm Ly}\alpha) = 8.7 L({\rm H}\alpha) 
\end{equation}
from Case B recombination theory (Brocklehurst 1971),
we can estimate the star formation rate using the Ly$\alpha$ luminosity;

\begin{equation}
SFR({\rm Ly}\alpha) = 9.1 \times 10^{-43} L({\rm Ly}\alpha) ~~~ 
M_\odot {\rm yr}^{-1}
\end{equation}
(see Hu et al. 1998). 
The results are given in the last column of Table 7. 
We note that the star formation rate derived here is reduced 
by the cosmic transmission (see appendix). 
The star formation rates range from 4.7 to 9.4 $M_\odot$ yr$^{-1}$
with an average of 6.4$\pm$1.6 $M_\odot$ yr$^{-1}$.
Although these values are typical to the Lyman break galaxies
at $z \sim$ 3 -- 4 (e.g., Steidel et al. 1999 and references therein),
the number density of the strong Ly$\alpha$ emitters like our sources
is rather small (see Figure 23). 

We examine whether or not the SFR derived from the Ly$\alpha$ luminosity
is consistent with that derived from the UV continuum luminosity
for our sample. The observed $i^\prime$ magnitude can be converted to 
a UV continuum luminosity at $\lambda$ = 1600 \AA. Using the following
relation (Kennicutt 1998; see also Madau et al. 1998),

\begin{equation}
SFR({\rm UV}) = 1.4 \times 10^{-28}L_{\nu}
~~ M_{\odot} ~ {\rm yr}^{-1},
\end{equation}
where $L_\nu$ is in units of ergs s$^{-1}$ Hz$^{-1}$, 
we estimate the SFR based on the rest-frame UV ($\lambda$1600 \AA)
continuum luminosity for each object. The results are summarized 
in Table 8.
Then we compare the two SFRs, $SFR$(Ly$\alpha$) and $SFR$(UV), 
for each object in Figure 24. It is shown that the two SFRs 
appear consistent within a factor of 2.

It is interesting to estimate the contribution of the star formation 
in the six Ly$\alpha$ emitter candidates to the co-moving 
cosmic star formation density (e.g., Madau et al. 1996). 
Integrating the star formation rates given in Table 7,
we obtain the co-moving star formation density for our sources,
$\rho_{\rm SFR} \sim 5.3 \times 10^{-4} M_\odot$ yr$^{-1}$ Mpc$^{-3}$.
In this estimate, we adopt an Einstein-De Sitter cosmology with
$H_0 = 50$ km s$^{-1}$ Mpc$^{-1}$, following the manner of Madau et al.
(1996). 

In Figure 25, we compare this star formation rate density with those 
of previous studies compiled by Trentham, Blain, \& Goldader (1999).
We also show the results obtained by CH98 and K00.
As shown in this figure, the star formation density derived in this study
is much smaller than the previous estimates. However, note that 
any reddening correction is not made for our data point (e.g.,
Pettini et al. 1999).
It is also noted that we do not integrate the star formation density 
of Ly$\alpha$ emitters assuming a certain luminosity function from
a lower to a upper limit. The reason for this is that there is no
reliable luminosity function for high-$z$ Ly$\alpha$ emitters
(see Figure 23).
Therefore, future careful investigations
will be absolutely necessary to estimate a more reliable contribution
of Ly$\alpha$ emitters to the cosmic star formation density at high redshift.

\section{SUMMARY}

We have presented our optical intermediate-band
($\lambda_{\rm c}$ = 5736 \AA ~ and $\Delta\lambda$ = 280 \AA)
and multicolor observations of the Subaru/XMM-Newton Deep Field
obtained with Suprime-Cam on the
8.2 m Subaru telescope. All the data were obtained during
the guaranteed time observations of the Suprime-Cam instrument.
The intermediate-band image covered a sky area with
10\minpoint62 $\times 12\minpoint40 \approx 132$ arcmin$^2$
in the Subaru/XMM-Newton Deep Field (Ouchi et al. 2001).
Our survey volume amounts to 93,952 $h_{0.7}^{-3}$ Mpc$^3$
when we adopt a flat universe with $\Omega_{\rm matter} = 0.3$, 
$\Omega_{\Lambda} = 0.7$
and $h=0.7$ where $h = H_0/($100 km s$^{-1}$ Mpc$^{-1}$).
We give a summary of our results below. 

(1) In our survey, we have found 23 emission-line sources whose
observed emission-line equivalent widths are greater than 250 \AA.
Their optical multicolor properties indicate 
six and two marginal emission-line sources are 
Ly$\alpha$ emitters at $z \approx$ 3.7 ($\Delta z \approx 0.22$)
They are either intense starburst
galaxies or active galactic nuclei like quasars at $z \approx$ 3.7.
Among the remaining 15 emission-line objects, 8 objects
may be either [O {\sc ii}]
$\lambda$3727 emitters at $z \approx 0.54$, H$\beta$
at $z \approx 0.18$, or
[O {\sc iii}]$\lambda$5007 emitters at $z \approx 0.15$.
The remaining 7 objects have been found only in the IA574 image.

(2) For the six Ly$\alpha$ emitters at $z \approx$ 3.7, we obtain
the average emission-line equivalent width of 
$<EW_0$(Ly$\alpha)> \simeq 103$ \AA ~ $\pm$ 53 \AA.
Their star formation rates range from 4.7 to 9.4 $M_\odot$ yr$^{-1}$
with an average of 6.4$\pm$1.6 $M_\odot$ yr$^{-1}$.
Although these values are typical to those of the Lyman break galaxies
at $z \sim$ 3 -- 4,
the number density of the strong Ly$\alpha$ emitters like our sources
appears rather small, 
since the present survey is not deep enough to detect faint emission 
line galaxies.

(3) We have estimated the contribution of the star formation
in the six Ly$\alpha$ emitter candidates to the co-moving
cosmic star formation density (e.g., Madau et al. 1996).
Integrating the star formation rates given in Table 6,
we obtain the co-moving star formation density for our sources,
$\rho_{\rm SFR} \sim 5.4 \times 10^{-4} M_\odot$ yr$^{-1}$ Mpc$^{-3}$.
In this estimate, we adopt an Einstein-De Sitter cosmology with
$H_0 = 50$ km s$^{-1}$ Mpc$^{-1}$, following the manner of Madau et al.
(1996).

\vspace{0.5cm}

We would like to thank the Subaru Telescope staff
for their invaluable help in commissioning the Suprime-Cam
that made these difficult observations possible.
Y. Shioya, M. Ouchi, H. Furusawa, and F. Nakata acknowledge
support from the Japan Society for the
Promotion of Science (JSPS) through JSPS Research Fellowships
for Young Scientists.
This work was financially supported in part by
the Ministry of Education, Science, Culture, and Technology
(Nos. 10044052, and 10304013).

\begin{center}
APPENDIX
\end{center}

\section*{The effect of cosmic transmission on the evaluation of Ly$\alpha$ 
emission }

We demonstrate here how the value of $\min (VR-IA574,R-IA574)$ 
is affected by the absorption of neutral hydrogen gas clouds 
between the object and us, so-called the cosmic transmission. 
We also show how the cosmic transmission affects the detectability of 
Ly$\alpha$ emitter candidates. 
In the text (section 4.2), we evaluate the value of $EW_{\rm obs}$ simply from 
$\min (VR-IA574,R-IA574)$ and the value of $EW_{\rm 0}$ by dividing 
$EW_{\rm obs}$ by $(1 + z_{\rm em})$, where $z_{\rm em}$ is a redshift of 
Ly$\alpha$ emitter candidate. 
Because of the cosmic transmission, the emission with wavelength of 
$\lambda < (1 + z_{\rm em}) \lambda_{\rm Ly \alpha}$ is dimmed 
as $F_{\rm obs}=F_{\rm int} \exp(-\tau_{\rm eff})$ where $F_{\rm obs}$ is 
the observed flux, $F_{\rm int}$ is the intrinsic flux, $\exp(-\tau_{\rm eff})$ is 
the cosmic transmission, and $\tau_{\rm eff}$ is the effective optical depth. 
We have simulated the effect of cosmic transmission on 
$\min (VR-IA574,R-IA574)$ and $EW_0$ as the following way. 
For this simulation we prepare the SED with Ly$\alpha$ emission by adding 
the emission line flux corresponding to $EW_0^{\rm model}$ with the Gaussian profile to the 
synthesized SED of young starburst galaxies, which is derived 
for the constant star formation rate with age of $10^8$ yr. 
We adopt the effective optical depth ($\tau_{\rm eff}$) formulated by 
Madau et al. (1996). 
Results are shown in Fig.24. 
If the cosmic transmission is 1 ($\tau_{\rm eff}=0$), the $\min (VR-IA574,R-IA574)$ of 
Ly$\alpha$ emitter with $E_0^{\rm obs}=100$ \AA~ is nearly constant 
and larger than 0.7 for the redshift between 3.61 and 3.82 
(see a long dashed line in Fig.24). 
On the other hand, adopting the average cosmic transmission, 
the value of $\min (VR-IA574,R-IA574)$ of Ly$\alpha$ emitter 
with $E_0^{\rm obs}=100$ \AA~ (a solid line) decreases with redshift 
and becomes smaller than 0.7 for redshift higher than 3.65. 
The $\min (VR-IA574,R-IA574)$ of Ly$\alpha$ emitters with $E_0^{\rm model} = 200$ \AA~ 
is always larger than 0.7 for redshift between 3.61 and 3.82 
even if the effect of the cosmic transmission is taking into account (dotted line). 
These imply that because of the cosmic transmission, the detectability of 
Ly$\alpha$ emitter depends on the redshift of the galaxy especially for 
Ly$\alpha$ emitters with smaller $EW_0({\rm Ly \alpha})$. 
We also show the $EW_0^{\rm obs}$ calculated simply from the value of 
$\min (VR-IA574,R-IA574)$ in the lower panel of Fig.24. 
Because of the cosmic transmission, the value of $EW_0^{\rm obs}$ is 
smaller than $EW_0^{\rm model}$. 
Taking account of this result, the star formation rate estimated in the text 
(section 4.3) may be considered as a lower limit of the star formation rate.



\clearpage

\begin{figure}
\begin{center}
\begin{tabular}{|c|}
\hline
fujita\_fig1.jpg \\
\hline
\end{tabular}
\end{center}
\caption{The transmission curves of the filters used in our observations.
\label{fig1}}
\end{figure}

\begin{figure}
\epsscale{0.4}
\begin{center}
\begin{tabular}{|c|}
\hline
fujita\_fig2.jpg \\
\hline
\end{tabular}
\end{center}
\caption{The Subaru/XMM-Newton field covered by the broad-band filters
         is shown by the large box (Ouchi et al. 2001). The sky area covered
         with the IA574 filter is shown by the small box. The background image 
         is taken from the Palomar Observatory Sky Survey Digital Atlas.
\label{fig2}}
\end{figure}

\begin{figure}
\begin{center}
\begin{tabular}{|c|}
\hline
fujita\_fig3.jpg \\
\hline
\end{tabular}
\end{center}
\caption{The final IA574 image (left panel) and the color image created by all the
         broad-band images (right panel).
\label{fig3}}
\end{figure}

\begin{figure}
\begin{center}
\begin{tabular}{|c|}
\hline
fujita\_fig4.jpg \\
\hline
\end{tabular}
\end{center}
\caption{The number counts of sources found in our $B$, $V$, $IA574$, 
         $R$, and $i^{\prime}$ images. The ordinates for $B$, $V$, 
         $R$, and $i^{\prime}$ are multiplied by $10^{-4}$, $10^{-2}$, 
         $10^2$, and $10^4$, respectively.
\label{fig4}}
\end{figure}

\begin{figure}
\begin{center}
\begin{tabular}{|c|}
\hline
fujita\_fig5.jpg \\
\hline
\end{tabular}
\end{center}
\caption{Detection completeness as a function of the apparent magnitude 
derived from the simulation (see text). Each symbol represents the 
adopted profiles of the artificial galaxies; de Vaucouleurs profile 
(cross) and exponential profile (plus). 
\label{fig5}}
\end{figure}

\begin{figure}
\begin{center}
\begin{tabular}{|c|}
\hline
fujita\_fig6.jpg \\
\hline
\end{tabular}
\end{center}
\caption{Distribution of the $VR-IA574$ color as a function of 
IA574 magnitude derived from the simulation (see text).
\label{fig6}}
\end{figure}

\begin{figure}
\begin{center}
\begin{tabular}{|c|}
\hline
fujita\_fig7.jpg \\
\hline
\end{tabular}
\end{center}
\caption{Objects detected to the apparent magnitude limit of $IA574=26$
         in the IA574-selected catalog.
         The horizontal broken line corresponds to the color of
         $VR-IA574=0.7$. Objects above this line have strong emission lines
         with $EW_{\rm obs} = 250$\AA ~ or greater.
         Solid lines and dotted lines show the distribution of $2 \sigma$ 
         and $3 \sigma$ error, respectively. 
\label{fig7}}
\end{figure}

\begin{figure}
\epsscale{0.30}
\begin{center}
\begin{tabular}{|c|}
\hline
fujita\_fig8.jpg \\
\hline
\end{tabular}
\end{center}
\caption{Plot of $VR-IA574$ versus $R-IA574$
         for the 81 objects found with $VR-IA574 > 0.7$ and 
         $IA574 < 25.4$ selection.
\label{fig8}}
\end{figure}

\begin{figure}
\begin{center}
\begin{tabular}{|c|}
\hline
fujita\_fig9.jpg \\
\hline
\end{tabular}
\end{center}
\caption{The diagram of $B-R$ color as a function of redshift for galaxies 
         with spectral energy distributions (SEDs) typical to
         E, Sbc, Scd, and Irr (CWW). The bluest is young starburst model (see text). 
         Redshift ranges for Ly$\alpha$ emitters at
         $z \approx 3.7$, [O II]$\lambda$3727 emitters at
         $z \approx 0.53$, and [O III]$\lambda$5007 emitters
         at $z \approx 0.15$ are also shown by shaded strips.
\label{fig9}}
\end{figure}

\begin{figure}
\begin{center}
\begin{tabular}{|c|}
\hline
fujita\_fig10.jpg \\
\hline
\end{tabular}
\end{center}
\caption{Diagram of objects in the IA574 selected catalog
         between $B-R$ and $R-i^\prime$ (right panel). Candidates of 
         Ly$\alpha$ emitters at $z \approx 3.7$ are located in the domain
         defined with the two color criteria; 1) $B-R \geq 1.0$, and 
         2) $R-i^\prime \leq 0.7$. Our final candidates of 
         Ly$\alpha$ emitters at $z \sim 3.7$ are shown by 
         red open circles. On the other hand, our final low-$z$ emitters 
         are shown by blue open squares which have $B-R < 1.0$.
         The marginal Ly$\alpha$ emitter candidates are shown by 
         pink open diamonds.  
         In the left panel, we show the model predictions for 
         the mean cosmic transmission (solid lines, see text). 
         For the model of young starburst galaxies, 
         we also show the case that the value of the effective optical depth is 
         a half of the mean value (dashed line). 
         The loci of model galaxies in the redshift between 3.6 and 3.83 
         are shown in thick lines. 
         Open circles show model galaxies at $z=0$.  
\label{fig10}}
\end{figure}

\begin{figure}
\begin{center}
\begin{tabular}{|c|}
\hline
fujita\_fig11.jpg \\
\hline
\end{tabular}
\end{center}
\caption{The broad-band and IA574 images of the most probable
         candidates of Ly$\alpha$ emitters at $z \approx 3.7$. 
         Each box is $16^{\prime \prime}$ on a side. 
         Each circle is $4^{\prime \prime}$ radius. 
         The SED is 
         also shown in right panel for each object.
\label{fig11}}
\end{figure}

\begin{figure}
\begin{center}
\begin{tabular}{|c|}
\hline
fujita\_fig12.jpg \\
\hline
\end{tabular}
\end{center}
\caption{The broad-band and IA574 images of emitter candidates 
         classified as `marginal' (see text). 
         Each box is $16^{\prime \prime}$ on a side. 
         Each circle is $4^{\prime \prime}$ radius. 
         The SED is 
         also shown in right panel for each object. 
\label{fig12}}
\end{figure}

\begin{figure}
\begin{center}
\begin{tabular}{|c|}
\hline
fujita\_fig13.jpg \\
\hline
\end{tabular}
\end{center}
\caption{The broad-band and IA574 images of the most probable
         candidates of Ly$\alpha$ emitters at lower redshifts. 
         Each box is $16^{\prime \prime}$ on a side. 
         Each circle is $4^{\prime \prime}$ radius. 
         The SED
         is also shown in right panel for each object. 
\label{fig13}}
\end{figure}

\begin{figure}
\begin{center}
\begin{tabular}{|c|}
\hline
fujita\_fig14.jpg \\
\hline
\end{tabular}
\end{center}
\caption{The broad-band and IA574 images of the unclassified 
         emission-line source which is detected only in the IA574 image.
         Each box is $16^{\prime \prime}$ on a side. 
         Each circle is $4^{\prime \prime}$ radius. 
         Its SED is also shown in right panel. 
\label{fig14}}
\end{figure}

\begin{figure}
\begin{center}
\begin{tabular}{|c|}
\hline
fujita\_fig15.jpg \\
\hline
\end{tabular}
\end{center}
\caption{Diagram of log EW([O II]) as a function of $B-R$.
         The data of the low-z emitter candidates in our survey are 
         shown by crosses.
         The observational data of nearby galaxies (dots) are taken from 
         Jansen et al. (2000).
         Solid lines show our model predictions, using 
         the population synthesis model GISSEL96 (Bruzual \& Charlot 1993) 
         with the star-formation time scale of 1 Gyr 
         and ages of 10 (reddest), 7, 4, 3, 2, 1, 0.5, 0.1, and 0.01 (bluest) 
         Gyr. We assume that $L({\rm H}\beta) =
         4.76 \times 10^{-13}N({\rm H}^0) ({\rm ergs \; s^{-1}})$, where 
         $N({\rm H}^0)$ is a ionizing photon production rate 
         in units of s$^{-1}$, 
         and $\rm log [O II]/H \beta = 0.0$ (the lower solid line) 
         and $0.5$ (the upper solid line). When emission-line flux is
         also taken into account in $R$ magnitude, the solid lines are
         shifted to the dashed lines.      
         Dotted vertical line shows a typical color of Irr galaxy.  
\label{fig15}}
\end{figure}

\begin{figure}
\begin{center}
\begin{tabular}{|c|}
\hline
fujita\_fig16.jpg \\
\hline
\end{tabular}
\end{center}
\caption{Diagram of log EW([O III]) as a function of $B-R$.
         The data of the low-z emitter candidates in our survey are 
         shown by crosses.
         The observational data (dots) are taken from Jansen et al. (2000).
         Dotted vertical line shows a typical color of Irr galaxy. 
         The meanings of solid and dashed lines are the same as those in
         Figure 15.
\label{fig16}}
\end{figure}

\begin{figure}
\begin{center}
\begin{tabular}{|c|}
\hline
fujita\_fig17.jpg \\
\hline
\end{tabular}
\end{center}
\caption{Comparison of histograms of $EW_{\rm obs}$(Ly$\alpha$) between
         the 6 Ly$\alpha$ emitter candidates at $z \approx 3.7$ (upper panel) 
         and 8 low-$z$ emitters (upper-middle panel), 
         2 marginal emitter candidates (lower-middle panel) 
         and 7 unclassified emitter candidates (lower panel).
\label{fig17}}
\end{figure}


\begin{figure}
\begin{center}
\begin{tabular}{|c|}
\hline
fujita\_fig18.jpg \\
\hline
\end{tabular}
\end{center}
\caption{A diagram between $EW_{\rm obs}$ and $VR$
         for the 8 low-$z$ emitter candidates.
\label{fig18}}
\end{figure}

\clearpage

\begin{figure}
\begin{center}
\begin{tabular}{|c|}
\hline
fujita\_fig19.jpg \\
\hline
\end{tabular}
\end{center}
\caption{The celestial positions of the 6 Ly$\alpha$ emitter candidates
         at $z \approx 3.7$ (open circles), marginal emitter candidates 
         (open diamonds), 8 low-$z$ emitter candidates (open squares), 
         and 7 unclassified emitter candidates (crosses).
\label{fig19}}
\end{figure}

\begin{figure}
\begin{center}
\begin{tabular}{|c|}
\hline
fujita\_fig20.jpg \\
\hline
\end{tabular}
\end{center}
\caption{Frequency distribution of $EW_0$(Ly$\alpha$) for the 6 
         Ly$\alpha$ emitter candidates at $z \approx 3.7$ (upper panel),
         the 2 marginal emitter candidates (middle), and 
         the 7 unclassified emitter candidates (lower panel).
\label{fig20}}
\end{figure}


\begin{figure}
\begin{center}
\begin{tabular}{|c|}
\hline
fujita\_fig21.jpg \\
\hline
\end{tabular}
\end{center}
\caption{Comparison of frequency distributions of $EW_0$(Ly$\alpha$)
         between our survey and those by CH98, S00, and MR01.
         The ordinate is the number of Ly$\alpha$ emitters per our survey 
         volume per interval of $EW_0$ = 50\AA. 
         Hatched region in out survey represents the marginal emitter candidates. 
\label{fig21}}
\end{figure}

\begin{figure}
\begin{center}
\begin{tabular}{|c|}
\hline
fujita\_fig22.jpg \\
\hline
\end{tabular}
\end{center}
\caption{Diagram between $EW_0$(Ly$\alpha$) and $VR$ magnitude
         for the 6 Ly$\alpha$ emitter candidates at $z \approx 3.7$ (filled circles)
         and 2 marginal emitter candidates (open diamonds). 
\label{fig22}}
\end{figure}

\begin{figure}
\begin{center}
\begin{tabular}{|c|}
\hline
fujita\_fig23.jpg \\
\hline
\end{tabular}
\end{center}
\caption{Distribution of Ly$\alpha$ 
         luminosity is compared to those derived by CH98 and K00. 
\label{fig23}}
\end{figure}

\begin{figure}
\begin{center}
\begin{tabular}{|c|}
\hline
fujita\_fig24.jpg \\
\hline
\end{tabular}
\end{center}
\caption{Comparison between $SFR$(Ly$\alpha$) and $SFR$(UV)
         for the 6 Ly$\alpha$ emitters at $z \approx$ 3.7. 
\label{fig24}}
\end{figure}

\begin{figure}
\begin{center}
\begin{tabular}{|c|}
\hline
fujita\_fig25.jpg \\
\hline
\end{tabular}
\end{center}
\caption{The star formation rate density at $z \approx 3.7$ derived
         in this study (diamond)
         is shown together with the previous investigations compiled
         by Trentham et al. (1999).  The data sources are Gallego et al. 
         (1996 - filled triangle), Treyer et al. (1998 - open triangle), 
         Tresse \& Maddox (1998 - open circle), Lilly et al. 
         (1996 - stars), Hammer \& Flores (1999 - open hexagons), 
         Connolly et al. (1997 - filled squares), Madau et al.
         (1996 - filled circles), and Pettini et al. (1999 - open squares).
\label{fig25}}
\end{figure}

\begin{figure}
\begin{center}
\begin{tabular}{|c|}
\hline
fujita\_fig26.jpg \\
\hline
\end{tabular}
\end{center}
\caption{The predicted value of $\min (VR-IA574,R-IA574)$ (upper panel) 
         and $E_0^{\rm obs}({\rm Ly \alpha})$, which is simply evaluated 
         from $\min (VR-IA574,R-IA574)$, as a function of 
         $z$ and $EW_0^{\rm model}({\rm Ly \alpha})$. 
         Solid (dotted, short-dashed) line shows a model galaxy with 
         $EW_0^{\rm model}({\rm Ly \alpha})=100$ (200, 400) \AA~ affected by the cosmic transmission. 
         We also show a model galaxy with $EW_0^{\rm model}({\rm Ly \alpha})=100$ \AA~ 
         and effective opacity is zero as a long dashed line. 
\label{fig26}}
\end{figure}

\clearpage

\begin{deluxetable}{lccc}
\tablenum{1}
\tablecaption{A journal of observations}
\tablewidth{0pt}
\tablehead{
\colhead{Band} & 
\colhead{Obs. Date} & 
\colhead{Total Integ. Time (s)}  &
\colhead{$m_{\rm lim}$(AB)\tablenotemark{a}} 
}
\startdata
$B$     & 2000 November 24, 25 & 10620 & 27.6 \\
$V$     & 2000 November 26, 27 &  4800 & 26.4 \\
$R$     & 2000 August 1 &  2520 & 26.3  \\
        & 2000 November 21 - 24  & 3480  & 26.5 \\
        & Total                  & 6000  & 26.7 \\
$i^\prime$ & 2000 November 25  & 2700 & 26.2 \\
$IA574$  & 2000 August 4 &  3600 & 25.7  \\
\enddata
\tablenotetext{a}{The limiting magnitude (3$\sigma$).}
\end{deluxetable}


{\tiny
\begin{deluxetable}{rccccccccccccc}
\tablenum{2}
\tablecaption{Photometric properties of Ly$\alpha$ emitter candidates at $z \approx 3.7$}
\tablewidth{0pt}
\tablehead{
\colhead{No.} &
\colhead{$\alpha$(J2000)} &
\colhead{$\delta$(J2000)}  &
\colhead{$EW_{\rm obs}$} &
\colhead{$B$} &
\colhead{$V$} &
\colhead{$IA574$} &
\colhead{$R$} &
\colhead{$i'$} &
\colhead{$VR$} &
\colhead{$VR-IA574$} &
\colhead{$R-IA574$} &
\colhead{$B-R$} &
\colhead{$R-i^\prime$} \\
\colhead{} &
\colhead{h ~~ m ~~ s} &
\colhead{$^\circ$ ~~ $^\prime$ ~~ $^{\prime\prime}$}  &
\colhead{(\AA)} & & & & & & & & & & 
}
\startdata
1 & 2 17 54.90 & $-$5 09 13.7 & 1014 $\pm$  439 &     27.73 &     26.35 &    24.88 &     26.54 &     26.16 & $>$ 27.02 & $>$ 2.14 &     1.66 &     1.19 &     0.38 \\
2 & 2 17 34.30 & $-$5 09 35.9 &  454 $\pm$  273 &     27.46 &     26.54 &    25.37 &     26.42 &     26.64 &     26.49 &     1.11 &     1.05 &     1.04 &  $-$0.22 \\
3 & 2 17 37.87 & $-$5 00 11.7 &  439 $\pm$  220 &     28.04 &     26.45 &    25.18 &     26.20 &     25.75 &     26.35 &     1.18 &     1.02 &     1.84 &     0.45 \\
4 & 2 17 46.26 & $-$5 11 42.5 &  414 $\pm$  159 &     27.71 &     25.76 &    24.83 &     26.25 &     25.74 &     25.81 &     0.99 &     1.42 &     1.46 &     0.51 \\
5 & 2 17 38.09 & $-$5 11 20.0 &  307 $\pm$  149 &     27.16 &     25.84 &    25.03 &     25.92 &     25.71 &     25.84 &     0.80 &     0.89 &     1.24 &     0.21 \\
6 & 2 18 06.71 & $-$5 10 05.5 &  269 $\pm$  115 &     27.34 &     25.88 &    24.88 &     25.61 &     25.66 &     25.79 &     0.91 &     0.73 &     1.73 &  $-$0.05 \\
\enddata
\end{deluxetable}

\newpage
\begin{deluxetable}{rccccccccccccc}
\tablenum{3}
\tablecaption{Photometric properties of marginal Ly$\alpha$ emitter candidates}
\tablewidth{0pt}
\tablehead{
\colhead{No.} &
\colhead{$\alpha$(J2000)} &
\colhead{$\delta$(J2000)}  &
\colhead{$EW_{\rm obs}$} &
\colhead{$B$} &
\colhead{$V$} &
\colhead{$IA574$} &
\colhead{$R$} &
\colhead{$i'$} &
\colhead{$VR$} &
\colhead{$VR-IA574$} &
\colhead{$R-IA574$} &
\colhead{$B-R$} &
\colhead{$R-i^\prime$} \\
\colhead{} &
\colhead{h ~~ m ~~ s} &
\colhead{$^\circ$ ~~ $^\prime$ ~~ $^{\prime\prime}$}  &
\colhead{(\AA)} & & & & & & & & & & 
}
\startdata
1 & 2 17 43.48 & $-$5 07 02.9 & 1022 $\pm$  635 &     27.51 &     26.84 &    25.27 &     26.94 &     26.64 & $>$ 27.02 & $>$ 1.75 &     1.67 &     0.57 &     0.30 \\
2 & 2 17 54.10 & $-$5 07 54.5 &  687 $\pm$  287 &     27.56 &     26.17 &    24.90 &     26.74 & $>$ 26.64 &     26.24 &     1.35 &     1.84 &     0.83 & $<$ 0.10 \\
\enddata
\end{deluxetable}

\begin{deluxetable}{rccccccccccccc}
\tablenum{4}
\tablecaption{Photometric properties of low-$z$ emitter candidates}
\tablewidth{0pt}
\tablehead{
\colhead{No.} &
\colhead{$\alpha$(J2000)} &
\colhead{$\delta$(J2000)}  &
\colhead{$EW_{\rm obs}$} &
\colhead{$B$} &
\colhead{$V$} &
\colhead{$IA574$} &
\colhead{$R$} &
\colhead{$i'$} &
\colhead{$VR$} &
\colhead{$VR-IA574$} &
\colhead{$R-IA574$} &
\colhead{$B-R$} &
\colhead{$R-i^\prime$} \\
\colhead{} &
\colhead{h ~~ m ~~ s} &
\colhead{$^\circ$ ~~ $^\prime$ ~~ $^{\prime\prime}$}  &
\colhead{(\AA)} & & & & & & & & & & 
}
\startdata
1 & 2 17 25.93 & $-$5 04 30.4 &      $>$  1177  &     27.75 & $>$ 26.84 &    25.23 & $>$ 27.12 & $>$ 26.64 & $>$ 27.02 & $>$ 1.79 & $>$ 1.89 & $<$ 0.63 & \nodata  \\
2 & 2 17 57.87 & $-$5 11 43.5 &      $>$  1091  &     27.52 & $>$ 26.84 &    25.30 & $>$ 27.12 & $>$ 26.64 & $>$ 27.02 & $>$ 1.72 & $>$ 1.82 & $<$ 0.40 & \nodata  \\
3 & 2 17 51.99 & $-$5 09 05.0 &  536 $\pm$  244 &     26.51 &     26.57 &    25.07 &     26.24 &     26.51 &     26.43 &     1.36 &     1.16 &     0.28 &  $-$0.27 \\
4 & 2 17 51.77 & $-$5 08 01.8 &  535 $\pm$  261 &     26.62 &     26.41 &    25.15 &     26.31 &     26.14 &     26.35 &     1.20 &     1.16 &     0.30 &     0.17 \\
5 & 2 17 38.24 & $-$5 09 14.8 &  284 $\pm$  173 &     26.37 & $>$ 26.84 &    25.29 &     26.05 &     26.50 & $>$ 27.02 & $>$ 1.73 &     0.76 &     0.32 &  $-$0.45 \\
6 & 2 17 54.42 & $-$5 08 23.9 &  268 $\pm$  105 &     25.75 &     25.50 &    24.75 &     25.49 &     25.76 &     25.48 &     0.73 &     0.74 &     0.26 &  $-$0.27 \\
7 & 2 17 28.61 & $-$5 08 54.1 &  261 $\pm$  154 &     26.53 &     25.96 &    25.19 &     25.94 &     26.17 &     25.91 &     0.71 &     0.75 &     0.59 &  $-$0.23 \\
8 & 2 17 51.73 & $-$5 00 54.9 &  254 $\pm$  153 &     26.15 &     26.32 &    25.24 &     25.94 &     26.08 &     26.22 &     0.98 &     0.70 &     0.21 &  $-$0.14 \\
\enddata
\end{deluxetable}

\begin{deluxetable}{rccccccccccccc}
\tablenum{5}
\tablecaption{Photometric properties of unclassified emitter candidates}
\tablewidth{0pt}
\tablehead{
\colhead{No.} &
\colhead{$\alpha$(J2000)} &
\colhead{$\delta$(J2000)}  &
\colhead{$EW_{\rm obs}$} &
\colhead{$B$} &
\colhead{$V$} &
\colhead{$IA574$} &
\colhead{$R$} &
\colhead{$i'$} &
\colhead{$VR$} &
\colhead{$VR-IA574$} &
\colhead{$R-IA574$} &
\colhead{$B-R$} &
\colhead{$R-i^\prime$} \\
\colhead{} &
\colhead{h ~~ m ~~ s} &
\colhead{$^\circ$ ~~ $^\prime$ ~~ $^{\prime\prime}$}  &
\colhead{(\AA)} & & & & & & & & & & 
}
\startdata
1 & 2 17 43.08 & $-$5 07 01.2 &      $>$  1508  & $>$ 28.04 & $>$ 26.84 &    25.01 & $>$ 27.12 & $>$ 26.64 & $>$ 27.02 & $>$ 2.01 & $>$ 2.11 & \nodata  & \nodata  \\
2 & 2 17 39.25 & $-$5 08 27.3 &      $>$  1199  & $>$ 28.04 & $>$ 26.84 &    25.21 & $>$ 27.12 & $>$ 26.64 & $>$ 27.02 & $>$ 1.81 & $>$ 1.91 & \nodata  & \nodata  \\
3 & 2 17 57.10 & $-$5 12 22.2 &      $>$  1092  & $>$ 28.04 & $>$ 26.84 &    25.30 & $>$ 27.12 & $>$ 26.64 & $>$ 27.02 & $>$ 1.73 & $>$ 1.82 & \nodata  & \nodata  \\
4 & 2 17 26.82 & $-$5 01 12.4 &      $>$  1080  & $>$ 28.04 & $>$ 26.84 &    25.31 & $>$ 27.12 & $>$ 26.64 & $>$ 27.02 & $>$ 1.72 & $>$ 1.81 & \nodata  & \nodata  \\
5 & 2 17 43.20 & $-$5 07 02.3 &      $>$  1074  & $>$ 28.04 & $>$ 26.84 &    25.31 & $>$ 27.12 & $>$ 26.64 & $>$ 27.02 & $>$ 1.71 & $>$ 1.81 & \nodata  & \nodata  \\
6 & 2 17 52.84 & $-$5 09 50.4 &      $>$  1008  & $>$ 28.04 & $>$ 26.84 &    25.36 & $>$ 27.12 & $>$ 26.64 & $>$ 27.02 & $>$ 1.66 & $>$ 1.75 & \nodata  & \nodata  \\
7 & 2 17 42.91 & $-$5 11 09.3 &  569 $\pm$  332 & $>$ 28.04 & $>$ 26.84 &    25.35 &     26.55 & $>$ 26.64 & $>$ 27.02 & $>$ 1.67 &     1.20 & $>$ 1.49 & $<-$0.09 \\
\enddata
\end{deluxetable}


{
\scriptsize
\begin{deluxetable}{lcccccccc}
\tablenum{6}
\tablewidth{7.in}
\tablecaption{%
A summary of the Ly$\alpha$-emitter surveys
}

\tablehead{
   \colhead{Survey\tablenotemark{a}} &
   \colhead{Field\tablenotemark{b}} &
   \colhead{Field Type\tablenotemark{c}} &
   \colhead{$z_{\rm c}$\tablenotemark{d}} &
   \colhead{$(z_{\rm min}, z_{\rm max})$\tablenotemark{e}} &
   \colhead{$V$\tablenotemark{f}} &
   \colhead{$EW_{\rm lim}({\rm Ly}\alpha)$\tablenotemark{g}} &
   \colhead{$N({\rm Ly}\alpha)$\tablenotemark{h}} &
   \colhead{$n({\rm Ly}\alpha)$\tablenotemark{i}}
}
\startdata
CH98 & HDF & B& 3.4 & (3.41, 3.47) & 5205 & 115 & 5 & $9.6 \times 10^{-4}$ \nl
CH98 & SSA22 & B& 3.4 & (3.41, 3.47) & 5205 & 90 & 7 & $1.3 \times 10^{-3}$  \nl
\hline
K99 & 53W002 & T& 2.4 & (2.32, 2.45) & 85338 & 92 & 19 & $2.2 \times 10^{-4}$ \nl
K99 & HU Aqr & B& 2.4 & (2.32, 2.45) & 85338 & 241 & 1 & $1.2 \times 10^{-5}$ \nl  
K99 & NGC 6251 & B& 2.4 & (2.32, 2.45) & 85338 & \nodata & 0 & 0 \nl
K99 & 53W002E & T& 2.55 & (2.49, 2.61) & 78588 & 291 & 1 & $1.3 \times 10^{-5}$ \nl
K99 & 53W002N & T& 2.55 & (2.49, 2.61) & 78588 & 155 & 1 & $1.3 \times 10^{-5}$ \nl
K99 & 53W002NE & T& 2.55 & (2.49, 2.61) & 78588 & 184 & 4 & $5.1 \times 10^{-5}$ \nl
\hline
S00 & LBGS\tablenotemark{j} & B& 3.09 & (3.07, 3.12) & 16741 & 80 & 72
 & $4.3 \times 10^{-3}$ \nl
\hline
K00 & Virgo\tablenotemark{k} & B& 3.14 & (3.12, 3.15) & 6020 & \nodata &
8 & $1.3 \times 10^{-3}$ \nl 
\hline
This study & Subaru/XMM & B& 3.72 & (3.60, 3.83) &  93952 & 254 & 6 
&  $6.4 \times 10^{-5}$ \nl
\enddata
\tablenotetext{a}{CH98 = Cowie \& Hu (1998), K99 = Keel et al. (1999),
     S00 = Steidel et al. (2000), and K00 = Kudritzki et al. (2000).}
\tablenotetext{b}{The name of the field.} 
\tablenotetext{c}{The field type; B=blank field, and T=targeted field.} 
\tablenotetext{d}{The central redshift corresponding to the central
     wavelength of the narrow-band filter ($\lambda_{\rm c}$).}
\tablenotetext{e}{The minimum and maximum redshift covered by the
      narrow-band filter.}
\tablenotetext{f}{The co-moving volume covered by the survey in
      units of $h_{0.7}^{-3}$  Mpc$^3$ with $\Omega_{\rm matter} = 0.3$
      and $\Omega_\Lambda = 0.7$.}
\tablenotetext{g}{The smallest equivalent width of the Ly$\alpha$ emission
      detected in the survey in units of \AA ~ in the observed frame.}
\tablenotetext{h}{The number of Ly$\alpha$ emitters found in the survey.}
\tablenotetext{i}{The number density of Ly$\alpha$ emitters found in the survey 
      in units of $h_{0.7}^3$ Mpc$^{-3}$.}
\tablenotetext{j}{The LBG (Lyman break galaxies) spike region.}
\tablenotetext{k}{La Palma field (Mendez et al. 1997) in the Virgo cluster.}
\end{deluxetable}
}
}
\begin{deluxetable}{rccc}
\tablenum{7}
\tablecaption{Ly$\alpha$ luminosity and star formation rate for the
Ly$\alpha$ emitter candidates at $z \approx 3.7$}
\tablewidth{0pt}
\tablehead{
\colhead{No.} &
\colhead{$EW_0$} &
\colhead{$L({\rm Ly}\alpha)$} &
\colhead{$SFR$(Ly$\alpha$)}  \\
\colhead{} &
\colhead{(\AA)} &
\colhead{($h_{0.7}^{-2}$ ergs s$^{-1}$)} &
\colhead{($h_{0.7}^{-2} M_\odot$ yr$^{-1}$)} 
}
\startdata
 1 & 216  & $1.04 \times 10^{43}$ & 9.4   \\
 2 &  97  & $5.20 \times 10^{42}$ & 4.7   \\
 3 &  93  & $6.14 \times 10^{42}$ & 5.6   \\
 4 &  88  & $8.27 \times 10^{42}$ & 7.5   \\
 5 &  65  & $6.02 \times 10^{42}$ & 5.5   \\
 6 &  57  & $6.48 \times 10^{42}$ & 5.9   \\
\enddata
\end{deluxetable}

\begin{deluxetable}{rcccc}
\tablenum{8}
\tablecaption{UV continuum luminosity and star formation rate for the
Ly$\alpha$ emitter candidates at $z \approx 3.7$}
\tablewidth{0pt}
\tablehead{
\colhead{No.} &
\colhead{$i^\prime$} &
\colhead{$f_\nu$($i^\prime$)} &
\colhead{$L_{1600}$\tablenotemark{a}} &
\colhead{$SFR$(UV)} \\
\colhead{} &
\colhead{(AB mag)} &
\colhead{($10^{-29}$ ergs s$^{-1}$ cm$^{-2}$ Hz$^{-1}$)} &
\colhead{($10^{29}$ ergs s$^{-1}$ Hz$^{-1}$)} &
\colhead{($h_{0.7}^{-2} M_\odot$ yr$^{-1}$)}
}
\startdata
 1 & 26.16   & 0.125  & 0.339  & 4.75 \\
 2 &$>26.20$ &$<0.120$&$<0.327$& $<4.58$ \\
 3 & 25.75   & 0.182  & 0.495  & 6.93 \\
 4 & 25.74   & 0.184  & 0.500  & 6.99 \\
 5 & 25.71   & 0.189  & 0.514  & 7.19 \\
 6 & 25.66   & 0.198  & 0.538  & 7.53 \\
\enddata
\tablenotetext{a}{The UV continuum luminosity at $\lambda$=1600 \AA.}
\end{deluxetable}

\end{document}